\documentclass[superscriptaddress,pra,preprint]{revtex4-1}
\usepackage{amsfonts}
\usepackage{amssymb}
\usepackage[dvips]{graphicx}
\usepackage{multirow}
\usepackage{amsmath}

\newcommand{\e}[2][-\imath]{e^{#1 #2}}
\newcommand{\ii}{\imath}
\newcommand{\shalf}{\frac{1}{\sqrt{2}}}
\newcommand{\half}{\frac{1}{2}}

\newcommand{\f}[1]{\textrm{#1}}
\newcommand{\B}[1]{\mathbf{#1}}
\newcommand{\eq}[1]{\begin{equation}#1\end{equation}}
\newcommand{\eqa}[1]{ \begin{eqnarray}#1\end{eqnarray}}
\renewcommand{\vector}[4][r]{
\left(\begin{array}{#1}
#2 \\
#3 \\
#4 \\
\end{array}\right)}

\begin{document}

\title{A class of compact entities in three component Bose - Einstein condensates. }

\author{Piotr Sza\'{n}kowski}
\affiliation{
Institute of Theoretical Physics, University of Warsaw, ul. Ho\.{z}a 69, PL--00--681 Warszawa, Poland
}
\author{Marek Trippenbach}
\affiliation{
Institute of Theoretical Physics, University of Warsaw, ul. Ho\.{z}a 69, PL--00--681 Warszawa, Poland
}
\author{Eryk Infeld}
\affiliation{So\l{}tan Institute for Nuclear Studies, ul. Ho\.{z}a 69, PL-00-681 Warsaw, Poland}
\author{George Rowlands}
\affiliation{Department of Physics, University of Warwick, Coventry CV4 7AL, England}

\begin{abstract}
We introduce a new class of soliton-like entities in spinor three component BECs. These entities generalize well known solitons.
For special values of coupling constants, the system considered is {\it Completely Integrable} and supports $N$ soliton solutions.
The one-soliton solutions can be generalized to systems with different values of coupling constants. However, they no longer interact
elastically. When two so generalized solitons collide, a spin component oscillation is observed in both emerging entities. We propose to call these
newly found entities {\it oscillatons}. They propagate without dispersion and retain their character after collisions. We derived an exact mathematical model
for oscillatons and showed that the well known one soliton solutions are a particular case.

\end{abstract}
\maketitle

\section{Introduction}

The idea of spinor condensates was first suggested in seminal papers of Ho
\cite{Shenoy+Ho:BinaryMixturesOfBEC} and Ohmi
\cite{Machida:BECWithInternalDoF,Machida:SpinDomainInBEC}. The experimental
creation of spinor condensates \cite{Ketterle:OpticalConfinementOfBEC}, in
which the spin degree of freedom, frozen in magnetic traps, comes into play,
opened the possibility to observe phenomena that are not present in single
component Bose Einstein condensates. These include the formation of spin
domains \cite{Ketterle:SpinDomainsInFroundStateOfBEC} and spin textures
\cite{Stamper-Kurn:SpinTextures}. A theoretical description of the formation of
spin domains can be found, for example in \cite{Tsubota:DomainFormationInBEC}.
A spinor condensate formed by atoms with spin $F$ is described by a macroscopic
wave function with $2F+1$ components. Here we focus on the $F=1$ case, which
has been studied in a number of theoretical works. The ground state structure
was investigated by several authors, for instance in
\cite{Ho:PhaseDiagramOfSpinorBEC,Ho:GroundStateOfSpinorBEC,Ueda:ExactEigenstatesOfSpinorBEC}.
Even multicomponent vector solitons with $F=1$ have been predicted; bright
solitons in \cite{Wadati:SpinorSolitons}, dark solitons
\cite{Wadati:DarkSolitons}, as well as gap solitons
\cite{Malomed:SolitonsNonlinearModulation} (the latter type requires the
presence of an optical lattice).

We investigate the dynamics of an $F = 1$ spinor Bose Einstein condensate for a
wide range of scattering length. In particular, we address the general problem
of spin soliton collisions. For one specific ratio of the scattering lengths,
Wadati and coworkers in \cite{Wadati:BrightSpinorSolitons,Wadati:KdV} found a
complete classification of the one soliton solution with respect to the spin
states and even presented an explicit formula of the two-soliton solution. One
soliton solutions come in two classes: polar, and ferromagnetic solitons
\cite{Wadati:BrightSpinorSolitons,Wadati:KdV}. Both can be generalized to a
wider set of scattering lengths. Here we consider all possible values of the
ratio of scattering lengths. Our system is no longer integrable, but some of
the one soliton solutions can be generalized to obtain solutions that preserve
their shape throughout.

The paper is organized as follows: in chapter 2 we show that $F=1$ one soliton solutions can be generalized for arbitrary nonlinear coupling, we discuss their shape and collisions. In chapter 3 we introduce a new kind of soliton solutions, which we call oscillatons. We study their dynamics and interactions.


\section{General considerations}

In this section we review and generalize results concerning the system of spinor Gross-Pitaevskii equations for the case of $F=1$ and equal coupling constants. This system is completely integrable and was thoroughly investigated by Wadati, Ieda and Miyakawa~\cite{Wadati:BrightSpinorSolitons}. Here we concentrate on bright soliton solutions.

To begin with, we consider a dilute gas of trapped bosonic atoms with hyperfine spin $F = 1$. The wavefunction in vector form is $\B\Phi\left(x,t\right)=\left(\Phi_1,\Phi_0,\Phi_{-1}\right)^T$ and it must satisfy the spinor Gross Pitaevski equation
\eq{\label{eq:mainequation}
\ii\hbar\partial_t\B{\Phi}
=
\left[ -\frac{\hbar^2}{2 M}\partial_x^2
 + c_0\B{\Phi}^\dagger \B{\Phi} +
 c_2\left(
     \sum_{\alpha=1}^3\left(\B{\Phi}^\dagger\hat f^\alpha\B{\Phi}\right)\hat f^\alpha
\right)\right]\B{\Phi}.
}
Here $\hat f^\alpha (\alpha=1,2,3)$ are the angular momentum operators in 3x3 representation and $c_0$ is negative to allow for bright soliton formation.

In the paper of Ieda {\it et al} \cite{Wadati:BrightSpinorSolitons} the authors considered this system with coupling constants $c_2=c_0\equiv c <0$. In this case Eq.~(\ref{eq:mainequation}) describes a completely integrable system. The authors find $\f N$ soliton solutions via the Hirota method. In particular, they present both $\f{N}=1$ and $\f{N}=2$ solutions explicitly.

\subsection{One soliton solutions}

We introduce dimensionless units:
$x\rightarrow \frac{\hbar^2\,L^2}{N|c_0|}\,x$ and
$t\rightarrow \frac{2M}{\hbar}\left(\frac{\hbar^2\,L^2}{N|c_0|}\right)^2.$ Here $N$ is the number of atoms and $L$ is a characteristic length of the problem (e.g.~$L = \sigma_\perp/\sqrt{3\pi}$ where $\sigma_\perp$ is the transverse size of trap confining the semi-one dimensional condensate).  When we express Eq.~(\ref{eq:mainequation}) in these units, and
divide the equation by $|c_0|$, all the coefficients but the ratio between self and cross nonlinear coupling $\frac{-c_2}{|c_0|}$, which we denote by $\gamma$, will be equal to one. To allow for the formation of bright solitons $c_0$ must be negative. The dimensionless form of our equation is
\eq{\label{eq:MainEQ}
\imath\partial_t\B\Phi = \Big[-\partial_x^2 -\B\Phi^\dagger\B\Phi
-\gamma\left(\B\Phi^\dagger\hat f^\alpha\B\Phi\right)\hat f^\alpha\Big]\B\Phi.
}
When $\gamma=1$ the general one-soliton solution is given by
\eq{\label{eq:WadatiOneSoliton}
\B\Phi = \frac{2k\e[\imath]{\varphi}}{1+\frac{1}{4k^2}e^{-2z}+k^2|\chi^\dagger\bar\chi|^2e^{2z}}
\left[\frac{1}{2k}e^{-z}\chi+k(\chi^\dagger\bar\chi)^\ast e^z\bar\chi\right],
}
where $z=k(x-x_0-2pt)$ a coordinate for observing soliton's envelope moving
with velocity $2p,$ $\varphi=px+(k^2-p^2)t$ a coordinate for observing the soliton'
s carrier wave, $\chi=(\chi_{+1},\chi_0,\chi_{-1})^T$ -
the polarization of soliton (normalized spinor), and
$\bar\chi=(\chi_{-1}^\ast,-\chi_0^\ast,\chi_{+1}^\ast)^T$ - time reversed
polarization.

The solitons can be classified according to the value of the parameter
$\left|\chi^\dagger\bar\chi\right|=\left|\chi^\dagger\hat T\chi\right|\equiv\langle T\rangle$
the mean value of the time reversal operator. $\hat T$ is defined as $\hat T =
\e{\pi \hat f^y}\,\hat K,$ where $\hat K$ is a complex conjugate operator.
$\langle T\rangle$ can take values ranging from $0$ to $1.$ Solutions with
$\langle T\rangle$ taking extreme values are of the greatest interest to us,
because they can be generalized to systems with general $\gamma.$ Solitons with
intermediate values of $\langle T\rangle$ seem to be unique for
integrable systems. We distinguish three classes of solitons \\
{\it 1. Ferromagnetic state}\\
When $\langle T\rangle = 0$ Eq.(\ref{eq:WadatiOneSoliton}) simplifies into separable form
\eq{\label{eq:WadatiFerro}
\B\Phi = k\, \f{sech}\left[k(x-x_0'-2pt)\right]\e[\imath]{px}\e[\imath]{(k^2-p^2)t}\chi.
}
Furthermore, the condition $\chi^\dagger\bar\chi=0$ implies that $\chi$ can be written as
\eq{\label{eq:WadatiFerroSoliton}
\chi^{(\f{ferro})}=
 \e{(\theta-\tau)}
 \vector[c]
  {\e{\beta}\cos^2\frac{\alpha}{2}}
  {\sqrt 2\cos\frac{\alpha}{2}\sin\frac{\alpha}{2}}
  {\e[\imath]{\beta}\sin^2\frac{\alpha}{2}}=
 \e[\imath]{\tau}\hat{\mathcal U}(\beta,\alpha,\theta)\vector{1}{0}{0}.
}
This is the spin state which minimizes energy in a system of $c_2<0.$  This kind of solution can be generalized for any $\gamma\ne1$. We do so by replacing $k$ multiplying the $\f{sech}$ function in Eq.~\ref{eq:WadatiFerroSoliton} with an appropriate $\gamma$ dependent coefficient. One can check that appropriate solution has the form
\eq{\label{eq:GeneralizeFerroSoliton}
\B\Phi^{(\f{ferro})}=\sqrt{\frac{2}{1+\gamma}}\,k\,\f{sech}\left[k(x-x_0-2pt)\right]
 \e[\imath]{p\,x}\e[\imath]{\left(k^2-p^2\right)t}\,
\e[\imath]{\tau}\hat{\mathcal U}\,\vector{1}{0}{0}.}
We call it a {\it generalized ferromagnetic soliton}. The total number of atoms is
\eq{
N_{\f{tot}}=\int_{-\infty}^{\infty}dx\,\B\Phi^\dagger\B\Phi=
 \frac{4}{1+\gamma}\,k,
}
the total mean spin
\eq{
\B f_{\f{tot}}=\int_{-\infty}^\infty dx\,\hat{\B e}_\alpha(\B\Phi^\dagger \hat f^\alpha\B\Phi) =
 N_{\f{tot}}\vector[c]{\sin\alpha\cos\beta}{\sin\alpha\sin\beta}{\cos\alpha}.
}
Here $\hat{\B e}_\alpha$ are versors of the coordinate system ($\alpha=x,y,z,$ summation for repeating indices). Finally the total momentum and energy of the generalized ferromagnetic soliton are
{\setlength\arraycolsep{0.1em}\eqa{
P^{\f{ferro}}_{\f{tot}}&=&\int_{-\infty}^\infty dx\,\B\Phi^\dagger(-\imath\partial_x\B\Phi)=N_{\f{tot}}\,p,\\
E^{\f{ferro}}&=&\int_{-\infty}^\infty
dx\,\left[\partial_x\B\Phi^\dagger\partial_x\B\Phi-\half(\B\Phi^\dagger\B\Phi)^2
-\frac{\gamma}{2}(\B\Phi^\dagger\hat f^\alpha\B\Phi)(\B\Phi^\dagger\hat f^\alpha\B\Phi)\right]=\nonumber\\
&=&N_{\f{tot}}\left(p^2-\frac{N_{\f{tot}}^2(1+\gamma)^2}{48}\right).
}} \\
{\it 2. The polar state} \\
Considering Eq. (\ref{eq:WadatiOneSoliton})
$\langle T\rangle =1$, we recover a normal sech-type soliton:
\eq{\label{eq:WadatiPolar}
\B\Phi=\sqrt 2\,k\, \f{sech}\left[k(x-x_0-2pt)\right]\e[\imath]{px}\e[\imath]{(k^2-p^2)t}\chi.
}
The constrain $\chi^\dagger\bar\chi=1$ implies that
\eq{
\chi^{(\f{polar})} =
 \e[\imath]{\tau}
  \vector[c]
   {-\shalf\e{\beta}\sin\alpha}
   {\cos\alpha}
   {\shalf\e[\imath]{\beta}\sin\alpha} =
 \e[\imath]{\tau}\hat{\mathcal U}(\beta,\alpha,\theta)\vector{0}{1}{0},
}
and one can check that the local mean spin density vanishes identically. Generalization of this soliton solution is straightforward. Since it is a
spinless state, the spin mixing interaction term in Eq. (\ref{eq:MainEQ}) vanishes and what remains is stratified by (\ref{eq:WadatiPolar}) for all $\gamma.$
We will refer to this solution as a{\it generalized polar soliton}
\eq{\label{eq:GeneralizedPolarSoliton}
\B\Phi^{(\f{polar})} = \sqrt 2\,k\,\f{sech}\left[k(x-x_0-2pt)\right]
 \e[\imath]{p\,x}\e[\imath]{\left(k^2-p^2\right)t}
 \e[\imath]{\tau}\hat{\mathcal U}\vector{0}{1}{0}.
} Notice that the amplitude of the soliton is different from that of the
ferromagnetic soliton, which leads to the different relation between the total
number of atoms and parameter $k$ \eq{ N_{\f{tot}} = 4k. } The energy
difference between ferromagnetic and polar solitons, with the same number of
atoms, is: \eq{ E^{\f{ferro}}-E^{\f{polar}}=
-\frac{1}{48}N_{\f{tot}}^3\gamma(2+\gamma).
} \\
{\it 3. Split solitons}\\
If $0<\langle T\rangle <1,$ for every moment $t,$ each component of the local
mean spin density vector $f^\alpha (x,t)= \B\Phi^\dagger\hat f^\alpha\B\Phi,$
is an anisymmetric function of $x$ with respect to a certain point
$x_{\f{node}}(t).$

\begin{figure}[h]
\centering\includegraphics[scale=0.45]{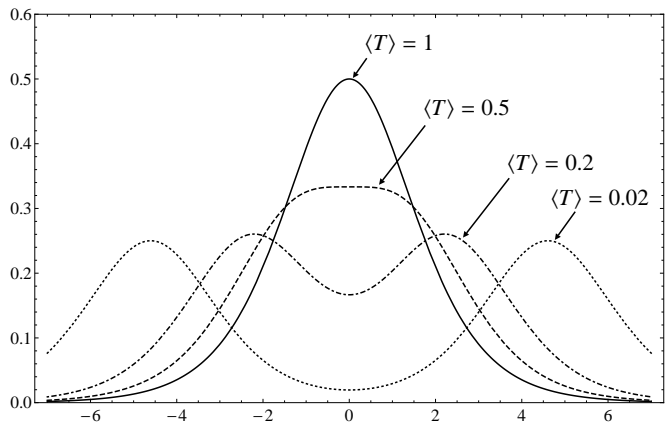}
\caption{Density profiles ($n(x)=\B\Phi^\dagger\B\Phi$) of split solitons for different values of $\langle T\rangle$, $\gamma=1$.}
\label{fig:splitsolitons}
\end{figure}

This implies that the total mean spin of this state, $\B f_{\f{tot}}=\int
dx\,\B f(x),$ is equal to $0.$ Careful examination of the density profile of
this kind of soliton reveals the reason for this. For $\langle T\rangle$ close
to $0$ the density splits into two disjointed peaks traveling with the same
velocity. Each of these peaks is actually a ferromagnetic soliton with mean
spins anti parallel to each other. As $\langle T\rangle$ approaches $1$ the
peaks begin to merge, consequently creating a single entity without spin - a
polar soliton. Figure (\ref{fig:splitsolitons}) shows the density profile
$n(x)=\B\Phi^\dagger\B\Phi,$ of split soliton for different values of $\langle
T\rangle$ and $\gamma=1$ from Eq.~(\ref{eq:WadatiOneSoliton}).

\subsection{Collisions}
\subsubsection{Elastic Collisions}

We begin with a short review of the integrable case. The only effect of a two soliton collision in the case of scalar solitons is a phase shift. The wave function after the collision can acquire additional phase and translation ($\B\Phi\rightarrow \e[\imath]{\tau}\B\Phi,$ and $\B\Phi(x,t)\rightarrow\B\Phi(x-\Delta x,t).$)~\cite{Ablowitz+Segur:Solitons+InverseScattering}.
In the context of the analysis of the previous section we can distinguish the following cases: \\
{\it (a) Polar-polar solitons collisions}
In a polar-polar soliton collision, the solitons emerge unaltered, aside from
phase changes. It is the result of the general rule: polar soliton cannot change
polarization of it's partner.\\
{\it (b) Ferromagnetic-ferromagnetic solitons collisions}
Ferromagnetic solitons change their phases and can rotate each other's
polarization. However, the values of $\langle T\rangle$ of each of
the solitons remain equal to $0$; in this kind of collision, solitons can't
change their type. \\
{\it (c) Polar-ferromagnetic solitons collision} In a collision, ferromagnetic
solitons experience only phase shifts; polarizations remain unchanged. This is
a consequence of the inability of a polar soliton to influence the
polarizations of other solitons when they interact. In the case of a polar
soliton, a combination of phase shifts and polarization rotation can change
$\langle T\rangle.$ This means that a polar soliton can be transformed into a
{\it split soliton} in the collision. However, $\langle T\rangle$ can never
reach $0,$ because the total spin must be conserved and there is no spin
transfer in the collision. A polar soliton will not change into a split soliton
if the polarizations of ferromagnetic and polar solitons are orthogonal:
$\chi^{(\f{polar})\dagger}\chi^{(\f{ferro})}=0.$

\subsubsection{Generalized soliton collisions}

Previously we have seen that two classes of one-soliton solutions can be
generalized to (almost) arbitrary $\gamma$. However, in order to call these
solutions real solitons one has to examine their mutual interactions. We have
conducted a series of numerical experiments on collisions of generalized
solitons for various values of $\gamma.$ We discovered that non-dissipating,
localized entities emerge in the wake of the collision. Although those entities
resemble solitons, there is a major difference: they are no longer stationary -
populations of magnetic components are oscillating with a well defined
frequency. This behavior is generic, it occurs for almost all values of
$\gamma$ (with exceptions of $\gamma=1$, when system is integrable and
$\gamma=0$, when there are no spin mixing interactions) and all configurations
of collisions. We propose to call these oscillating, soliton-like entities
oscillatons. The creation of oscillatons is indeed generic, however, depending
on the details of the collisions (such as the sign of $\gamma$ and classes of
solitons participating in it) this process can be accompanied by some side
effects (for example: a short period of intense radiation, a small momentum
transfer).

We consider a head-on collision of two generalized solitons for some $\gamma.$
At $t=0$, when solitons are far apart, the wave function is, within a good
approximation, a sum of two one-soliton wave functions $ \B\Phi(t=0) \approx
\B\Phi^{(1)} + \B\Phi^{(2)}$. The wave functions of solitons $\B\Phi^{(i)}$ are
given by (\ref{eq:GeneralizeFerroSoliton}) for generalized ferromagnetic
solitons and (\ref{eq:GeneralizedPolarSoliton}) for generalized polar solitons.

Each of the participating solitons is described by a set of parameters: gauge
phase $\tau$, Euler angles $\beta,\alpha$ and $\theta$ (in the case of a polar soliton
$\theta$ is a dummy variable), momentum $p$ and amplitude $k.$ In principle, the result
 of a collision can depend on all of these parameters. However, Galilean and
gauge-rotation invariance of the equation reduces the number of parameters
significantly. Firstly, the Galilean invariance allows us to fix one of the
solitons in place (we will call it {\it the target}) and set the other one in
motion (we will call it {\it the bullet}). Equivalently, the collision can be
viewed in the reference frame moving with the target soliton. Secondly,
gauge-rotation invariance allows us to fix the target's polarization to
``standard'' orientation ($(1,0,0)^T$ for the ferromagnetic and $(0,1,0)^T$ for
the polar target) and the gauge phase of the target can be set to zero. In this
work we restricted considerations to equal norms. This configuration allows for
observation of the behavior of a post-target oscillaton at large times. As an
example we will use a polar-ferromagnetic collision (for polar-polar collision see \cite{MyPRL:Oscillatons}).

\begin{figure}[h!]
\centering
\includegraphics[scale=.5]{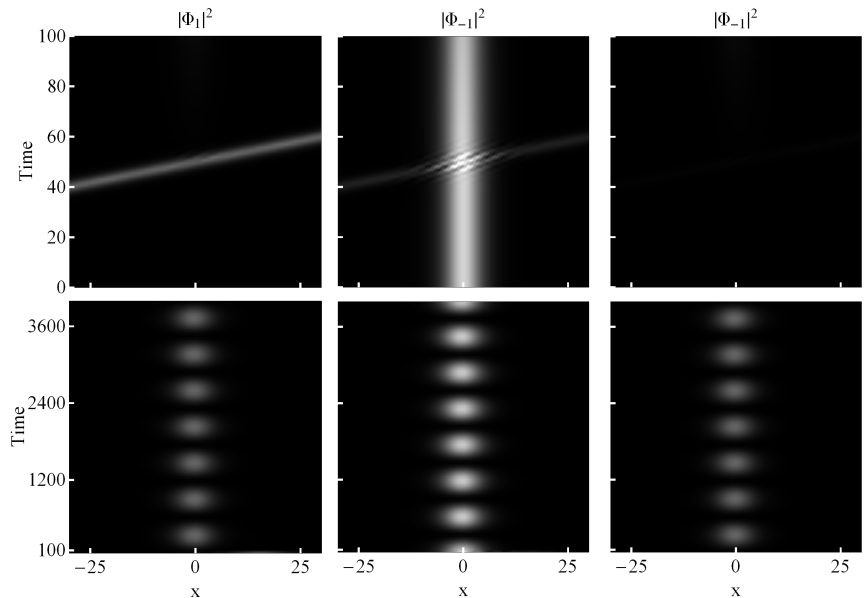}
\caption{Top: Density plot of $|\Phi_1|^2$ (left), $|\Phi_0|^2$ (center) and
$|\Phi_{-1}|^2$ (right) for a collision of a stationary polar and a moving
 ferromagnetic soliton. Here $\gamma=-1/3,$ the momentum is $p=1.5,$ and the Euler
angles are $\beta=\alpha=\pi/4,$ $\tau=\theta=0.$ Both solitons are normalized to
$1.$
Bottom: Details of the initially stationary polar soliton pictured after the
collision and to a much longer time scale. Observe the oscillatory
character of the wave function components.}
\label{fig:SpaceTimeCollision}
\end{figure}

Figures (\ref{fig:SpaceTimeCollision}) show space-time plots of collisions and
propagation of post-target oscillatons in a time scale much longer then the
time of collision. The oscillatory behavior can be clearly seen. Each of those
components can be fitted with the function $f_m(t) = a_m + b_m\,\cos(\omega t
+\varphi_m),$ where $m=1,0,-1.$ Relative phases between components are such,
that the total density
$|\Phi_{1}(x_{\f{max}})|^2+|\Phi_0(x_\f{max})|^2+|\Phi_{-1}(x_\f{max})|^2$ is
constant in time. The frequency $\omega$ defines the frequency of an
oscillaton.

\begin{figure}[h!]
\centering
\includegraphics[scale=.8]{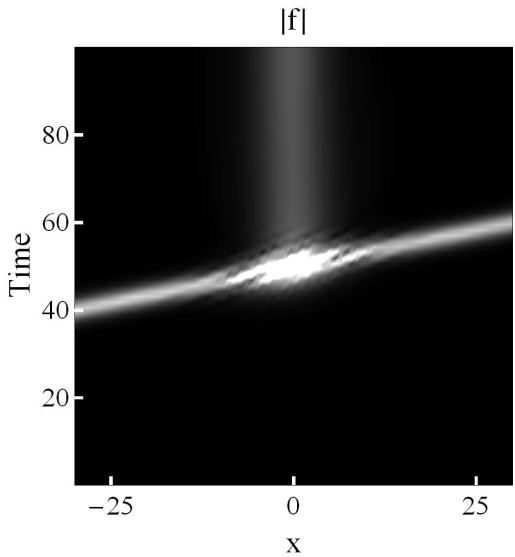}
\caption{Density plot of local spin density $|\B f(x,t)|$ for the collision
showed in Fig. (\ref{fig:SpaceTimeCollision}).} \label{fig:Spin}
\end{figure}

An important feature of the collision is the spin transfer.
Figures (\ref{fig:Spin}) show space-time density plot of local spin density
$|\B f(x,t)|=\sqrt{(\B\Phi^\dagger\hat f^\alpha\B\Phi)(\B\Phi^\dagger\hat f^\alpha\B\Phi)}.$
 At first, only a ferromagnetic soliton has spin. During the collision, a polar soliton, and hence a post-polar oscillaton, acquire some spin at the expense of
a post-ferro oscillaton. Spin densities of both oscillatons are constant
and the total spin ($|\B f_{\f{tot}}|=\int dx |\B f(x,t)|$) is conserved.
 In the case of polar-polar collision, both participants start with no spin.
During the collision, both target and bullet acquire spins, but the spin vectors
are anti parallel, so that the total spin is still $0.$

We found that the oscillation frequency is determined by the spin transferred
in the collision. It depends only on the magnitude of the spin vector, not its
direction: the greater the spin transfer, the greater the oscillation
frequency. The amount of spin transferred in the collision depends on all
collision parameters: the momentum $p$, Euler angles, norm and $\gamma.$

Dependance on momentum is mostly
due to the time of interaction; fast collision means less time for atoms to transfer
and weaker effects of interactions. We found that the frequency of oscillations
depends on the momentum as $\omega(p)\propto e^{-a\,p}/p^b,$ where $a$ and $b$ are
constants, fitted for a particular collision type. Notice that $\omega$ has a
maximum for some $p.$ The collisions are elastic (oscillatons are not
created) when spinors of participating solitons ($\chi$ in Eq.
(\ref{eq:GeneralizedPolarSoliton}) and Eq.
(\ref{eq:GeneralizeFerroSoliton})) are parallel or
perpendicular. In the case of parallel spinors, the equations effectively reduce to
the scalar Gross-Pitaevskii equation, which is integrable, hence, the collision is elastic.

Another noteworthy feature of the dynamics of the collisions is the very small
momentum transfer. Notice that despite a very long time of observation, the
oscillaton practically has not moved from its original position (aside from a
small ``recoil'', characteristic of polar-ferro soliton collision
~\cite{Wadati:BrightSpinorSolitons}. This means that the only effective way to
transfer energy between solitons/oscillatons is a transfer and redistribution
of atoms between magnetic components.

\begin{figure}[!h]
\centering
\includegraphics[scale=.6]{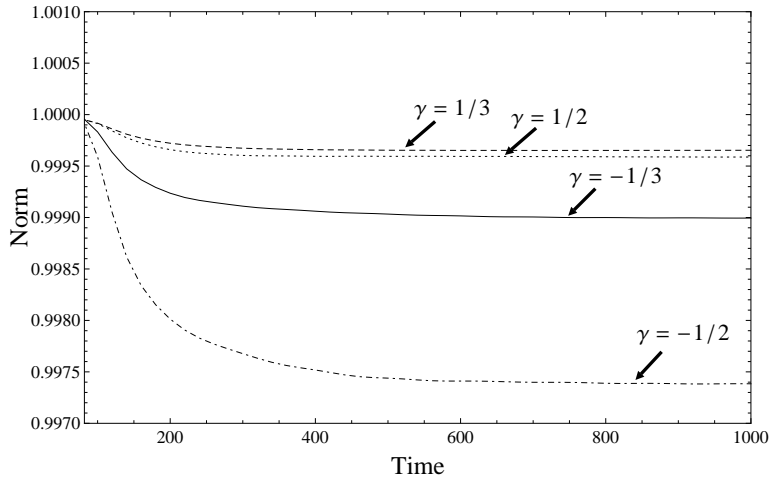}
\caption{Norm $\int \B\Phi^\dagger\B\Phi\,dx$ of the
Oscillatons created in collision of polar and
ferromagnetic soliton as a function of time and $\gamma.$ Collision setups are as in Fig. (\ref{fig:SpaceTimeCollision}). Atoms are lost due to radiation.}
\label{fig:Radiation}
\end{figure}

The collisions between generalized solitons are inelastic. As soon as
oscillatons split up, we observe a short period of intense radiation. The
actual intensity of this radiation strongly depends on the collision setup and
value of $\gamma.$ Figure (\ref{fig:Radiation}) shows norms of oscillaton as a
function of time.
Note the exponential-like norm decay to some fixed value, this atom loss is due
to radiation. In the collision solitons and/or oscillatons exchange atoms. This
leads to the spin and energy transfer. It seems that, in the collision,
solitons can acquire an extra amount of atoms and end up in a non stable state.
By releasing these excess particles, in the form of radiation, the oscillaton
may transfer into its equilibrium state. The frequency of oscillations is
insensitive to radiation.

\section{A new type of soliton: an oscillaton}

\subsection{One oscillaton solution}

In a collision of solitons, spin is transferred such that the spin vector of the outgoing oscillaton has a non-zero component in the $XY$ plane. The amplitude of oscillations can be affected by rotating the frame of reference, the frequency of oscillations however  does not change. Particularly, in a reference frames where the projection of the spin vector on the Z-axis was greater, the amplitude of oscillations dropped. In a special frame, in which spin vector is parallel to $Z$-axis, the amplitude is $0.$ On top of that, an oscillaton has a very simple form when viewed in this ``good'' reference frame: a $m=0$ component vanishes, $m=\pm 1$ have constant modulai with time dependence in the form of a linear phases increase. In summary, an oscillaton wave function can be modeled by the following anzats
\eq{\label{eq:OscillatonAnzats}
\B\Phi =
 \e[\imath]{\tau}\hat{\mathcal U}(\beta,\alpha,\theta)
 \left[
  \eta_{+}(x)\e[\imath]{\mu_{+}t}\vector{1}{0}{0} +
  \eta_{-}(x)\e[\imath]{\mu_{-}t}\vector{0}{0}{-1}
 \right],
}
where $\eta_\pm$ are real, symmetric functions, $\mu_\pm$ are positive, real numbers and $\hat{\mathcal U}$ is a spin rotation operator
parameterized by the Euler angles $\{\beta,\alpha,\theta\}.$ This anzats describes
oscillations of component populations, as can be seen by examining the modulai of components
\eqa{
|\Phi_0|^2 &=& \half\sin^2\alpha\left(\eta_+^2+\eta_-^2
  +2\eta_+\eta_-\cos\left[\omega\,t-2\theta\right]\right),\\
|\Phi_{\pm 1}|^2 &=& \eta_\pm^2\sin^4\frac{\alpha}{2}+\eta_\mp^2\cos^4\frac{\alpha}{2}
 -\half\eta_+\eta_-\sin^2\alpha\cos\left[\omega\,t-2\theta\right],
}
where $\omega\equiv\mu_+ - \mu_-.$ The local spin density in this state is
\eq{
\B f = \hat e_\alpha (\B\Phi^\dagger\hat f^\alpha\B\Phi) =
 \left(\eta_+^2 - \eta_-^2\right)
 \vector[c]
  {\sin\alpha\cos\beta}
  {\sin\alpha\sin\beta}
  {\cos\alpha}.
}
We find that the $\beta$ angle plays a role only in determining the orientation of the spin vector. We also find that the total spin is constant in time.

The total density profile of oscillaton is constant in time
\eq{
n = \B\Phi^\dagger\B\Phi = \eta_+^2+\eta_-^2.
}

This anzats substituted into Eq. (\ref{eq:MainEQ}) leads to the following
system of two coupled {\it ordinary} differential equations
\eq{\label{eq:OscillatonEquations} -\mu_\pm +
(1+\gamma)\eta_\pm^2+(1-\gamma)\eta_\mp^2+\frac{\eta_\pm ''}{\eta_\pm}=0. } The
problem has thus been reduced to solving ordinary differential equations.
Additionally we have a first integral \eq{ -\mu_+ \eta_+^2
-\mu_-\eta_-^2+\frac{1+\gamma}{2}\left(\eta_+^4+\eta_-^4\right)+(1-\gamma)\eta_+^2\eta_-^2+\eta_+'^2+\eta_-'^2
= \f{const} } The equations (\ref{eq:OscillatonEquations}) are nonlinear, so
amplitudes of wave function components $\eta_\pm$ determine chemical potentials
$\mu_\pm,$ and $\omega=\mu_+-\mu_-.$  For the case of $\mu_+=\mu_-\equiv\mu$
(or $\omega=0$) we have no oscillations. This implies
$\eta_+=\eta_-=\sqrt{\mu}\,\f{sech}(\sqrt{\mu}x)$ and the spinor part of the
wavefunction is proportional to $(1,0,-1)^T,$. This can be obtained from polar
spinor $(0,1,0)^T$ by a rotation $\hat{\mathcal U}.$ Hence that polar soliton
(\ref{eq:GeneralizedPolarSoliton}) is a special case of an oscillaton. When
$\mu_- = 0$ or $\mu_+ =0$ the ferromagnetic soliton
(\ref{eq:GeneralizeFerroSoliton}) is obtained from Eq.
(\ref{eq:OscillatonEquations}). Again, we see that a ferromagnetic soliton is
also a special case of an oscillaton.

This ansatz opens new possibilities of finding both exact and approximate
solutions. A particular case is when $\gamma=1$ (integrable system), for which
a solution for the pair of equations (\ref{eq:OscillatonEquations}) is obtained
explicitly. For this $\gamma$ the equations for $\eta_\pm$ decouple and each
can be solved analytically. The solutions of interest are \eq{ \eta_\pm(x) =
\sqrt{\mu_\pm}\,\f{sech}\left(\sqrt{\mu_\pm}\,x\right). } We have found a solution
that looks like being composed of two ferromagnetic solitons! As long as
$\mu_+\neq\mu_-$ there will be oscillations with frequency
$\omega=\mu_+-\mu_-,$ although, it is not possible to obtain such an oscillaton
in a collision.

In order to ascribe an oscillaton as obtained in a particular collision, one
has to follow the steps listed below.
\begin{enumerate}
\item Establish the orientation of the spin vector of the oscillaton and perform a
rotation to the reference frame in which this vector will be parallel to the
$Z$-axis We will call this frame of reference the {\it eigenframe}.
\item Determine the chemical potentials, the $\mu_\pm$ of $m=\pm 1$ components of the wave
function.
\item Insert the chemical potentials into the oscillaton equations (\ref{eq:OscillatonEquations})
and solve them for $\eta_+$ and $\eta_-$.
\end{enumerate}
The first two steps can always be completed. The third is problematic. In
general, the oscillaton equations (\ref{eq:OscillatonEquations}) are not
exactly solvable. However, in the following sections we will show that, in the
cases of oscillatons created in collisions of generalized polar and
ferromagnetic solitons, approximate solutions can indeed be found.

\subsubsection{Post-ferromagnetic oscillaton}

\begin{figure}[h!]
\centering
\includegraphics[scale=.6]{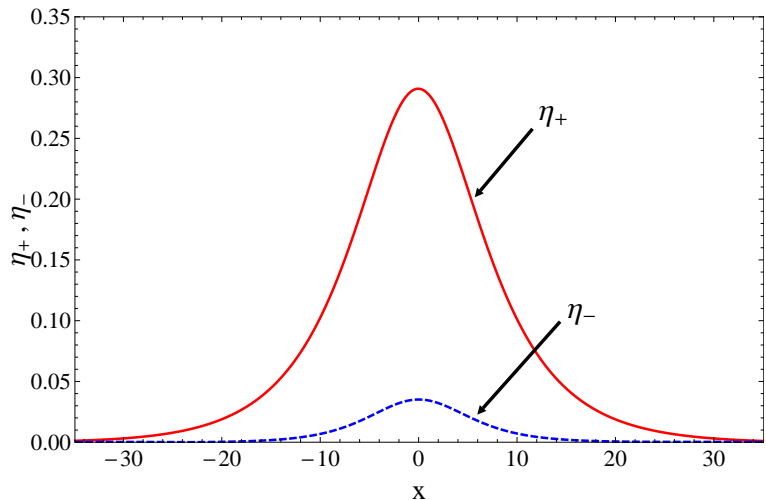}
\caption{(Color online) Modulus of components of post-ferromagnetic oscillaton viewed in its
eigenframe: $|\Phi_1|=\eta_+$ (red, solid line) and $|\Phi_{-1}|=\eta_-$ (blue
, dashed line). The oscillaton was created in a collision of the target
ferromagnetic soliton and the bullet polar soliton with $p=1.5,$ $\alpha=\beta=\pi/4,$ $\tau=\theta=0,$ and $\gamma=-1/3.$}
\label{fig:EigenPostFerroOscillaton}
\end{figure}

Figure (\ref{fig:EigenPostFerroOscillaton}) shows an oscillaton ``created''
out of a ferromagnetic soliton in polar-ferro collision, when viewed in its
eigenframe. In this collision $\gamma=-1/3,$ the ferromagnetic soliton was the
target and the polar soliton with $p=1.5,$ $\alpha=\beta=\pi/4,$
$\tau=\theta=0,$ was the bullet.

Looking back to Fig. (\ref{fig:EigenPostFerroOscillaton}), it is clear that one of the components of the wave function is much smaller then the other, or $\eta_+\gg \eta_-.$ In terms of our equations (\ref{eq:OscillatonEquations}) this means that we can neglect $\eta_-^2$ terms in comparison to $\eta_+^2$
\eqa{
-\mu_{+}+(1+\gamma)\eta_{+}^2+\frac{\eta_+ ''}{\eta_+} &\approx& 0
 \label{eq:FerroApprox1}\\
-\mu_- +(1-\gamma)\eta_{+}^2+\frac{\eta_- ''}{\eta_-} &\approx& 0.
 \label{eq:FerroApprox2}
} Now, the first equation depends only on $\eta_+$ and can be solved exactly:
$\eta_+ = \sqrt{2\mu_+/(1+\gamma)}\,\f{sech}\left(\sqrt{\mu_+}\,x\right).$ We next
insert this value into the equation for $\eta_-$, where it plays the role of a
trapping potential: \eq{\label{eq:Landau} \eta_- '' +\left[ -\mu_- + 2\mu_
 +\left(\frac{1-\gamma}{1+\gamma}\right)
 \f{sech}^2\left(\sqrt{\mu_+}\,x\right)\right]\eta_-=0.
}
Fortunately, this equation can be solved exactly~\cite{Landau+Lifshitz:vol3}. The even solution is
\eq{\label{eq:FerroOscilaton}
\eta_- \propto
 \f{sech}^{\epsilon}\left(\sqrt{\mu_+}\,x\right)\,
 P^{(\epsilon,\epsilon)}_{2n}\left[\f{tanh}\left(\sqrt{\mu_+}\,x\right)\right],
}
where $\epsilon = \sqrt{\mu_-/\mu_+},$ $n=0,1,2,\ldots$ and
$P^{(a,b)}_m\left[z\right]$ is a Jacobi polynomial, given by
\eq{
P^{(a,b)}_m\left[z\right]=\frac{(-1)^m}{2^m m!}
 (1+z)^{-a}(1-z)^{-b}\frac{d^m}{dz^m}
 (1+z)^{a+m}(1-z)^{b+m}.
}
The first two solutions are
\eqa{
\eta_-^{(n=0)} &\propto& \f{sech}^\epsilon \left(\sqrt{\mu_+}\,x\right), \nonumber\\
\eta_-^{(n=1)} &\propto& \f{sech}^\epsilon \left(\sqrt{\mu_+}\,x\right) \left[ 1
  - \left(\frac{1+2\epsilon}{2+2\epsilon}\right)\f{sech}^2\left(\sqrt{\mu_+}\,x\right)\right].
  \nonumber
}
Equation (\ref{eq:Landau}) imposes a quantization condition on the chemical potential
 $\mu_-$:
\eq{\label{eq:Spectrum}
\mu_- = \frac{\mu_+}{4}\left[ -1-4n+\sqrt{1+8\frac{1-\gamma}{1+\gamma}}\right]^2.
}
The condition $\mu_- \geq 0$ ($\eta_-$ must be a bounded state) gives the number of allowed energy levels $n.$ For $\gamma>-1/2$ there is only one eigenvalue,
corresponding to $n=0.$ As $\gamma$ approaches $-1/2$ from above the $n=1$ value appears. At $\gamma=-9/11,$ $n=2$ appears. As we near $\gamma=-1,$ the spectrum
becomes arbitrarily rich. Unfortunately, for $\gamma\rightarrow -1$ our approximation breaks down, as can be seen from Eq. (\ref{eq:FerroApprox1}). Luckily, from the physical point
of view, the most interesting cases are for $\gamma\sim 0,$ where the approximation should work.
\begin{figure}[h!]
\centering
\includegraphics[scale=.6]{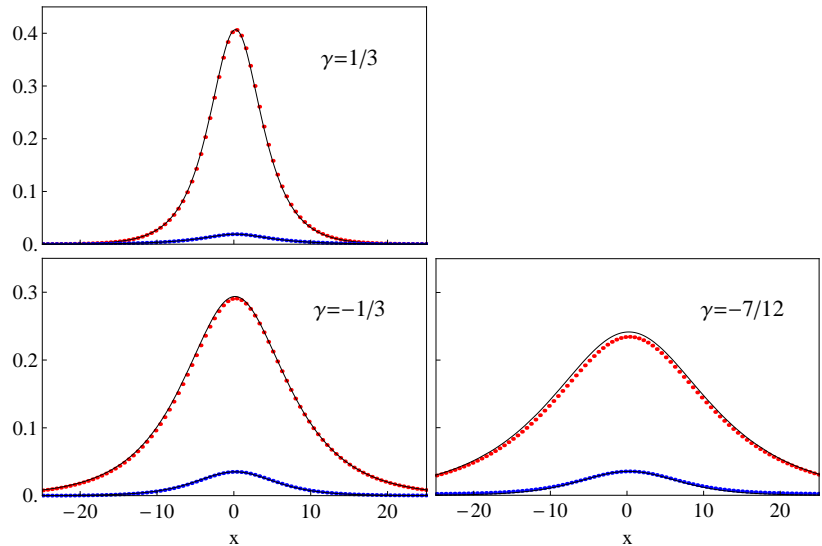}
\caption{Comparison of solutions of Eq.~(\ref{eq:FerroApprox1}) and
(\ref{eq:FerroApprox2}) with post-ferromagnetic oscillatons found numerically
viewed in their respective eigenframes. Oscillatons have been created in the
collision of a target ferromagnetic soliton and a bullet polar soliton with
$p=1.5,$ $\alpha=\beta=\pi/4,$ $\tau=\theta=0.$ The collisions were carried out
for three cases of $\gamma:$ $1/3$ (top-left panel), $-1/3$ (bottom-left panel)
and $-7/12$ (bottom-right panel). Agreement between the approximate solution
for $\eta_\pm$ (lines) and red/blue dots obtained in the simulation is very
good. In order to obtain the solutions the chemical potentials $\mu_\pm$ and
the normalization constant of $\eta_-$ had to be fitted to match the numerical
results. For all presented cases $n$ in Eq. (\ref{eq:Spectrum}) has been found
to be $0.$} \label{fig:PostFerroComp}
\end{figure}

Figure~(\ref{fig:PostFerroComp}) shows a comparison between solutions of
approximate equations~(\ref{eq:FerroApprox1}), (\ref{eq:FerroApprox2}) and
post-ferromagnetic oscillatons obtained in ferro-polar soliton collisions for
different values of $\gamma.$ In all cases the $\eta_-$ component ended up in
the ground state ($n=0$ in Eq.~(\ref{eq:Spectrum})), even for
$\gamma=-7/12<-1/2.$ Agreement between the model and the results of numerical
simulations is very good.

\subsubsection{Post-polar Oscillaton}

\begin{figure}[h!]
\centering
\includegraphics[scale=.6]{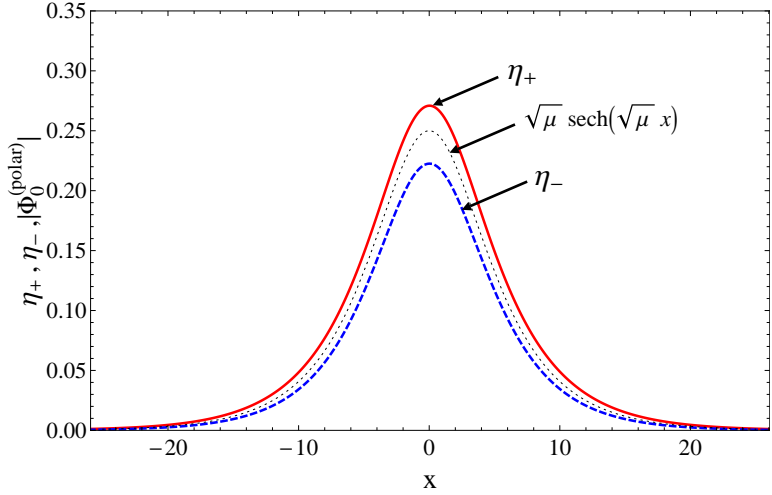}
\caption{(Color online) Modulus of components of post-polar oscillaton viewed in its
eigenframe: $|\Phi_1|=\eta_+$ (red,solid line) and $|\Phi_{-1}|=\eta_-$
(blue, dashed line). Components are compared with
$\sqrt{\mu}\,\f{sech}(\sqrt{\mu}\,x)$ (dotted line). The oscillaton was created
in the collision of the target polar
soliton and the bullet ferromagnetic soliton with $p=1.5,$ $\alpha=\beta=\pi/4,$
 $\tau=\theta=0,$ and $\gamma=-1/3.$}
\label{fig:EigenPostPolarOscillaton}
\end{figure}

Figure (\ref{fig:EigenPostPolarOscillaton}) compares components of the
wavefunction of a post-polar oscillaton viewed in its eigenframe and the
original polar soliton. Here we use as an example the collision setup presented
in Fig. (\ref{fig:SpaceTimeCollision}).

In this case, $\eta_+$ and $\eta_-$ are comparable. We propose the following ansatz:
\eq{\label{eq:PolarAnsatz}
\eta_\pm =\alpha_\pm\,\f{sech}^{\sqrt{\mu_\pm/\mu}}\left(\sqrt{\mu}\,x\right),
}
where $\mu$ is some ``central'' chemical potential. We anticipate that the central chemical potential is in fact the amplitude $\mu=k^2$ of the initial polar soliton (see Eq.~(\ref{eq:GeneralizedPolarSoliton})). This was confirmed by numerical simulations. Now, we will assume that the exponents in (\ref{eq:PolarAnsatz}) can be written in the following form
\eq{
\sqrt{\frac{\mu_\pm}{\mu}} \equiv \sqrt{1+\delta_\pm},
}
where $\delta_\pm$ are small corrections. Ansatz (\ref{eq:PolarAnsatz}) inserted into equations (\ref{eq:OscillatonEquations}) leads to the following
system
\eqa{\label{eq:PAnsatzEqn}
&&(1+\gamma)\alpha_\pm^2\f{sech}^{2\sqrt{1+\delta_\pm}}\left(\sqrt{\mu}\,x\right) +
\nonumber\\
&&+(1-\gamma)\alpha_\mp^2\f{sech}^{2\sqrt{1+\delta_\mp}}\left(\sqrt{\mu}\,x\right) +
\nonumber\\
&&-\mu_\pm\left(1+\frac{1}{\sqrt{1+\delta_\pm}}\right)\f{sech}^2\left(\sqrt{\mu}\,x\right)
=0. } In order to make further progress with equation (\ref{eq:PAnsatzEqn}) we
investigate the Taylor expansion of $\f{sech}^{2\sqrt{1+\delta}}$ in $\delta$
{\setlength\arraycolsep{0.1em} \eqa{\label{eq:TaylorExpansion}
\f{sech}^{2\sqrt{1+\delta}}\left(\sqrt{\mu}\,x\right) &=&
 \f{sech}^2 \left(\sqrt{\mu}\,x\right)
+ \left(\frac{d}{d\delta}\,\f{sech}^{2\sqrt{1+\delta}}\right)\Big|_{\delta=0}\,\delta
+ \ldots =\nonumber\\
&=& \f{sech}^2\left(\sqrt{\mu}\,x\right)\left[1 +
 \delta\,\ln\left(\f{sech}\left(\sqrt{\mu}\,x\right)\right)\right]+\ldots
} } For $|x|\lesssim 1/\sqrt{\mu},$ the second term in the expansion can be
neglected.  Within this region of $x,$ $\f{sech}^{2\sqrt{1+\delta}}\approx
\f{sech}^2$ and equation (\ref{eq:PAnsatzEqn}) can be satisfied, as long as
\eq{\label{eq:DeltaEqn} \mu_\pm+\sqrt{\mu}\,\sqrt{\mu_\pm}
=(1+\gamma)\alpha_\pm^2 +
 (1-\gamma)\alpha_\mp^2.
} (since $1+\delta_\pm = \mu_\pm/\mu$). This relation is valid as long as we
can approximate $\f{sech}^{2\sqrt{1+\delta}}$ by $\f{sech}^2$ in equation
(\ref{eq:PAnsatzEqn}). It is easy to see that this approximation will work for
$|x|\gg 1/\sqrt{\mu}.$ For large $x,$ $\f{sech}^{2\sqrt{\mu_\pm/\mu}}$ terms
can be dropped and Eq.~(\ref{eq:OscillatonEquations}) reduces to \eq{
-\mu_\pm+\frac{\eta_\pm ''}{\eta_\pm}=0, } satisfied by ansatz
(\ref{eq:PolarAnsatz}), as
$\alpha_\pm\f{sech}^{\sqrt{\mu_\pm/\mu}}(\sqrt{\mu}\,x) \rightarrow 2\alpha_\pm
\exp\left(-\sqrt{\mu} |x|\right),$ when $|x|\gg 1\/\sqrt{\mu}.$

Now we will establish validity of our approximation in the intermediate range
of $x,$ i.e. $|x|\gtrsim 1/\sqrt{\mu}.$ So far we haven't used the fact that
$\delta_\pm$ should be small. For small $\delta$ and $|x|\gtrsim 1\sqrt{\mu}$
the expansion (\ref{eq:TaylorExpansion}) can be written as
{\setlength\arraycolsep{0.1em} \eqa{
\f{sech}^{2\sqrt{1+\delta}}\left(\sqrt{\mu}\,x\right) &=&
 \f{sech}^2\left(\sqrt{\mu}\,x\right)\Bigg[
  1 + \delta\ln\left(\f{sech}\right) +. \nonumber\\
&&  +\delta^2\left(-\frac{1}{4}\ln(\f{sech})+\half\ln^2(\f{sech})\right)+\ldots\Bigg]
\nonumber\\
&\approx& \f{sech}^2(\sqrt{\mu}\,x)\left[ 1 + \delta\sqrt{\mu}\,|x|
 + \half \left(\delta\sqrt{\mu}\,|x|\right)^2 + \ldots \right] \nonumber\\
&=& \f{sech}^2\left(\sqrt{\mu}\,x\right)\,e^{\delta\sqrt{\mu}\,|x|}. }} This means,
that $\f{sech}^{2\sqrt{1+\delta}} \approx \f{sech}^2,$ if $|x| \ll
1/\delta\sqrt{\mu}$ and we conclude, that our approximation is valid everywhere
if $1\/\sqrt{\mu}\ll 1/\delta\sqrt{\mu},$ or $\delta \ll 1.$ In order to
convince ourself that $\delta_\pm$ are indeed small, we perform an estimate
using Eq. (\ref{eq:DeltaEqn}). As we saw in Fig.
(\ref{fig:EigenPostPolarOscillaton}) amplitudes of {\it oscillaton} components
differ by small corrections from the amplitude of the initial polar soliton.
Knowing this, we can write $\alpha_\pm^2 = \mu(1\pm\Delta)$, where $\mu$ is an
amplitude of the initial soliton and $\Delta$ is small. Inserting it into
equations (\ref{eq:DeltaEqn}) we get \eq{
\left(\sqrt{1+\delta_\pm}\right)^2+\sqrt{1+\delta_\pm} =
 2\left(1 \pm \gamma\,\Delta\right).
}
The solution to this equation is $\delta_\pm \approx \pm\frac{4}{3}\gamma\,\Delta$.

We see that, indeed, small differences in amplitudes implies small differences in chemical potentials. For an alternative calculations see Appendix \ref{apx:AppendixA}.

\begin{figure}[h!]
\centering
\includegraphics[scale=.6]{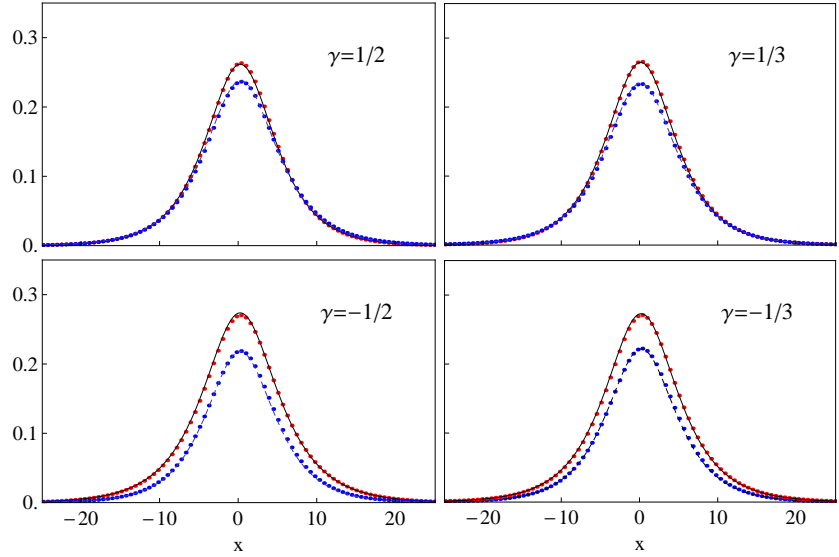}
\caption{Comparison of approximate solutions~(\ref{eq:PolarAnsatz}) and (\ref{eq:PAnsatzEqn}) with post-polar oscillatons found numerically
viewed in their respective eigenframes.
oscillatons have been created in a collision of the target polar
soliton and the bullet ferromagnetic soliton with $p=1.5,$ $\alpha=\beta=\pi/4,$
$\tau=\theta=0.$ The collisions were carried out for four cases of $\gamma:$
$1/2$ (top-left panel), $1/3$ (top-right panel), $-1/2$ (bottom-left panel) and
$-1/3$ (bottom-right panel). Agreement between approximate solution for
$\eta_\pm$ (solid/dashed line) and red/blue dots obtained in simulation is very
good. In order to obtain the solutions, chemical potentials $\mu_\pm$ had to be
fitted to mach the numerical results. The central chemical potential $\mu$ in
Eq.~(\ref{eq:PolarAnsatz}) and (\ref{eq:PAnsatzEqn}) is found to be the
amplitude of initial polar soliton $\mu=k^2$ (see
Eq.~(\ref{eq:GeneralizedPolarSoliton})).}
\label{fig:PostPolarComp}
\end{figure}
Figure (\ref{fig:PostPolarComp}) presents a comparison of approximate solutions~(\ref{eq:PolarAnsatz}) and (\ref{eq:PAnsatzEqn}) with
post-polar oscillatons obtained in polar-ferro solitons collisions for different values of $\gamma.$ The agreement between the model and numerical
results, for all $x,$ is very good.


\subsection{Oscillaton collisions}

The results of numerical experiments show that there is no qualitative difference between oscillatons and generalized solitons. We observe the same characteristic features: spin transfer, brief periods of radiation, small momentum transfer and so on. After the collision, oscillatons emerge altered, but nevertheless still described by our model. Besides a small atomic loss due to radiation and a tiny momentum change, oscillatons change their frequency of oscillations. Evidently, this is the result of
spin transfer during the collision - a mechanism we discussed above. The similarity between solitons and oscillatons collisions is no surprise. Our previous considerations showed that both polar and ferromagnetic solitons are indeed special cases of oscillatons - non oscillating ones.
In Table \ref{tab:one} we compare oscillaton parameters before and after a collision. We choose to look at changes in crucial parameters: chemical potentials $\mu_\pm$, total spin $\B f_{\f{tot}}$. We analyze four cases. First we consider post-polar vs post-polar oscillaton collision for (A) $\gamma= - 1/3$, and (B) $\gamma=1/3$.  We collide an oscillaton with small spin (I) with one of much larger spin (II). From the point of view of the oscillaton with greater spin, the collision was almost elastic; the relative change of chemical potentials $\mu_\pm$ and spins are very small. This behavior is somewhat similar to the elastic collision of two ferromagnetic solitons in the case of a completely integrable system
($\gamma=1$), when the solitons, due to interaction, only rotate each other's spins. On the other hand, the oscillaton with
smaller spin experiences not only a rotation of its spin vector, but also a substantial increase of its  magnitude.

Next we consider the collision of post-polar and post-ferromagnetic oscillatons for (C) $\gamma= -1/3$ and (D) $\gamma= 1/3$. The results are analogous. Post ferromagnetic oscillatons (labeled as II), with greater spin, collide almost elastically: chemical potentials and spins have hardly changed. The post-polar oscillatons (I), with smaller spin, similarly to the previous case, experiences not only reorientation of the spin vector, but also an increase of magnitude.
\begin{table}
\centering
\caption{Comparison of the oscillaton parameters before and after collision. Here we define
$|\B f_{\f{tot}}| \equiv \int|\B f^{(\f{before})}|\f d x$, 
$\Delta\mu_\pm \equiv \frac{\mu_\pm^{(\f{after})}-\mu_\pm^{(\f{before})}}{\mu_\pm^{(\f{before})}}$, 
$\Delta |\B f_{\f{tot}}| \equiv
\frac{\int |\B f^{(\f{after})}|-|\B f^{(\f{before})}|\f d x}
 {\int |\B f^{(\f{before})}|\f d x}$, 
$\cos\Delta\theta \equiv
\frac{\B f^{(\f{before})}\cdot\B f^{(\f{after})}}
{|\B f^{(\f{before})}||\B f^{(\f{after})}|}$
}

\begin{tabular}
{
|c|@{\quad}p{1.5cm}|@{\qquad}p{2cm}p{2cm}p{2cm}p{2cm}|}
\hline
&$|\B f_{\f{tot}}|$
&$\Delta\mu_+$
&$\Delta\mu_-$
&$\Delta|\B f_{\f{tot}}|$
&$\cos\Delta\theta$\\
\hline
\hline
A&\multicolumn{5}{l|}
{ Post---polar vs. post---polar oscillaton collision at $\gamma=-1/3$}\\
\hline
 I&0.05535&-0.09035&0.06098&3.325&0.9844\\
 II&0.2175&0.002253&-0.002147&-0.02487&-0.1902\\
\hline\hline
B&\multicolumn{5}{l|}{Post---polar vs. post---polar oscillaton collision at $\gamma=1/3$}\\
\hline
 I&0.0279&0.1021&-0.08473&6.947&0.822\\
 II&0.1106&-0.007527&0.004616&-0.1147&-0.733\\
\hline\hline
C&\multicolumn{5}{l|}{Post---polar vs. post---ferro oscillaton collision at $\gamma=-1/3$}\\
\hline
 I&0.05535&-0.06036&0.04176&-0.445&0.7197\\
 II&0.977&0.0004905&-0.0003409&0.001215&0.9944\\\hline\hline
D&\multicolumn{5}{l|}{ Post---polar vs. post---ferro oscillaton collision at $\gamma=1/3$}\\
\hline
 I&0.0279&0.01687&-0.01961&1.364&0.06856\\
 II&0.9939&-0.001359&0.000303&0.0009568&0.9994\\
\hline

\end{tabular}
\label{tab:one}
\end{table}

\section{Summary}

We considered a one dimensional, three component Bose-Einstein condensate with spin exchange interaction and general coupling constants $c_2$ and negative $c_0$.
The class of soliton-like solutions, universal to a wide range of coupling constants has been found. We called these solutions oscillatons. The mathematical model of a one-oscillaton solution have been derived.

Upon interacting with each other, oscillatons, similarly to solitons in an integrable system, retain theirs identities. However, unlike in
the soliton case, the collisions are not elastic.  Experimental realization of the ideas presented here was suggested earlier \cite{MyPRL:Oscillatons}.

\begin{acknowledgements}

The authors acknowledge support of a Polish Government Research Grant and also the Foundation for Polish Science Team Programme co-financed by the EU European Regional Development Fund.

\end{acknowledgements}


\appendix\section{Expansion in the case $\eta_+ \approx \eta_-$.}\label{apx:AppendixA}

Assume the difference between $\mu_+$ and $\mu_-$ to be small. Introduce $\mu_\pm = \mu(1\pm\delta)$ and $\xi = \sqrt \mu x.$ We now have $\eta_\pm = \sqrt{\frac{\mu}{1+\gamma}}\,f_\pm.$ Equations (\ref{eq:OscillatonEquations}) now lead to
\eq{\label{eq:AppendixA:feqn}
\frac{\f d^2 f_\pm}{\f d\xi^2} -f_\pm +f^3_\pm +\beta f^2_\mp\,f_\pm = \pm\delta f_\pm,
}
where $\beta = (1-\gamma)/(1+\gamma).$ We now expand $f_\pm$ in $\delta$:
\eq{
f_\pm \approx f^{(0)} + f_\pm^{(1)}\,\delta.
}
Equation (\ref{eq:AppendixA:feqn}) in zero order is:
\eq{
\begin{array}{lcr}
\frac{\f d^2\,f^{(0)}}{\f d \xi^2} - f^{(0)} +(1+\beta)f^{(0)^3}=0 &
\quad\Rightarrow\quad &
f^{(0)} = \sqrt{\frac{2}{1+\beta}}\,\f{sech}(\xi).\\
\end{array}
}
And in the next order we find
\eq{\label{eq:AppendixA:1stordereqns}
\hat L f^{(1)}_\pm + 2\beta f^{(0)^2} f_\mp^{(1)} = \pm f^{(0)},
}
where
\eq{\label{eq:AppendixA:L}
\hat L = \frac{\f d^2}{\f d \xi^2} -1 +(3+\beta)f^{(0)^2}=
 \frac{\f d^2}{\f d \xi^2} -1 + 2\left(\frac{3+\beta}{1+\beta}\right)\f{sech}^2\xi.
}
Adding the two equations (\ref{eq:AppendixA:1stordereqns}) yields
\eq{
\left[ \frac{\f d^2}{\f d \xi^2} -1 +3(1+\beta)f^{(0)^2} \right](f_+^{(1)}+f_-^{(1)}) =0
}
Solved by $f_+^{(1)}+f_-^{(1)} \propto d f^{(0)}(\xi)/{\f d\xi},$ corresponding to a shift in position, a trivial transformation. Thus, without loss of generality we may assume $f^{(1)}_+ = - f^{(1)}_-.$ Define
\eq{
f^{(1)}_+-f^{(1)}_- \equiv \sqrt{\frac{2}{1+\beta}}\,\Delta f.
}
We now have just one differential equation to solve:
\eq{\label{eq:AppendixA:deltafeqn}
\hat L \Delta f = \f{sech}(\xi).
}
The solution to the inhomogeneous equation is
$$
\Delta f_{\f{inh}}=\half\left(\frac{1+\beta}{3-\beta}\right)\cosh(\xi).
$$
This solution is ill behaved at large distances from the center. We must try to balance this by our choice of solution to the homogeneous equation. Write
\eq{
\begin{array}{lccr}
\hat L\Delta f_{\f{h}} = 0
&\; , \quad
& \Delta f_{\f{h}} = \f{sech}(\xi)\,F(z)
&\; ,\quad z \equiv \frac{e^{-\xi}}{e^\xi + e^{-\xi}}\\
\end{array}
}
Now
\eq{
z(1-z)\frac{\f d^2 F}{\f d z^2} + 2(1-2 z)\frac{\f d F}{\f d z} +
(\lambda - 2)F = 0,
}
where $\lambda = 2(3-\beta)/(1+\beta).$ This equation is solved by \cite{Whitaker+Watson:AcourseOfModernAnalysis}
\eq{
F(z) = A\:{}_2 F_1 \left[ \frac{\alpha}{2},\frac{\beta}{2},\half,(1-2z)^2 \right],
}
where ${}_2F_1$ is a hypergeometric function and
\eq{
\begin{array}{ccccc}
1-2z = \tanh\xi&
,\quad\alpha\beta = 2- \lambda&
,\quad\alpha = \frac{3+\kappa}{2}&
,\quad\beta = \frac{3-\kappa}{2}&
,\quad\kappa \equiv \sqrt{1+4\lambda}.\\
\end{array}\nonumber
}
We have taken the symmetric solution only. It is essential to demand that the hypergeometric function ${}_2F_1$ be non---polynomial, i.e. $\lambda$ not one of $2(n+1)(2n+1)$ ($n$ is an integer).

The full solution is now
\eq{
\Delta f = \frac{1}{\lambda}\cosh\xi + A\,\f{sech}\,\xi\:
 {}_2F_1\left[\frac{3+\kappa}{2},\frac{3-\kappa}{2},\half,\tanh^2\xi\right],
}
where $A$ is a constant to be determined. Now for $\xi\rightarrow\pm\infty,$ $\tanh^2\xi\rightarrow 1.$ We find that in this limit conveniently \cite{Morse+Feshbach:MethodsOfTheoreticalPhysics}
\eq{
\lim_{x\rightarrow 1}{}_2F_1\left[\frac{3+\kappa}{2},\frac{3-\kappa}{2},\half,x\right]
 = \frac{\Gamma\left(\half\right)\Gamma\left(1\right)}{\Gamma\left(\frac{3+\kappa}{2}\right)\Gamma\left(\frac{3-\kappa}{2}\right)}\,
 \frac{1}{1-x},
}
with $x = \tanh^2\xi,$ $1/(1-x) = \cosh^2\xi.$ Thus, for $\xi\rightarrow\pm\infty$
\eq{
\Delta f \rightarrow \frac{1}{\lambda}\cosh\xi +
 A\frac{\sqrt \pi}{\Gamma\left(\frac{3+\kappa}{2}\right)\Gamma\left(\frac{3-\kappa}{2}\right)}
 \cosh\xi
}
and so we choose $A= -\Gamma\left(\frac{3+\kappa}{2}\right)\Gamma\left(\frac{3-\kappa}{2}\right)/\sqrt \pi\lambda.$
\newline
The complete, well behaved solution is
\eq{
\Delta f = \frac{1}{\lambda}\left(\cosh\xi -
\frac{\Gamma\left(\frac{3+\kappa}{2}\right)\Gamma\left(\frac{3-\kappa}{2}\right)}{\sqrt \pi}
\f{sech}\,\xi\;{}_2F_1\left[\frac{3+\kappa}{2},\frac{3-\kappa}{2},\half,\tanh^2\xi\right]\right).
}
As indicated above, ${}_2F_1$ must not be polynomial. This is reflected in our solution, as the Gamma function of $-n$ is infinite. When ${}_2F_1$ is polynomial, e.g. $\f{sech}\,\xi$ for $\lambda = 2,$ $\f{sech}\,\xi(1-5\tanh^2\xi)$ for $\lambda = 12,$ it fails to balance the solution of the inhomogeneous equation at large distances. Our calculation is therefore somewhat flawed, though only for isolated points.

The case $\lambda=0$ must be considered separately. The solution to (\ref{eq:AppendixA:deltafeqn}) is, for $\beta=3$
\eq{
\Delta f = \xi\,\sinh\xi - \cosh\xi\ln\left(2\cosh\xi\right),
}
which is perfectly well behaved at large distances.


\begin{thebibliography}{99}
\bibitem{Shenoy+Ho:BinaryMixturesOfBEC}
  T.-L. Ho and V.~B. Shenoy, Phys. Rev. Lett., \textbf{77}, 3276 (1996)
\bibitem{Machida:BECWithInternalDoF}
  T. Ohmi and K. Machida, Journal of the Physical Society of Japan, \textbf{67}, 1822 (1998)
\bibitem{Machida:SpinDomainInBEC}%
  T. Isoshima, K. Machida, and T. Ohmi, {Phys. Rev. A}, \textbf{60}, {4857} ({1999})%
\bibitem{Ketterle:OpticalConfinementOfBEC}%
  {D.~M.} {Stamper-Kurn}, {M.~R.} {Andrews}, {A.~P.} {Chikkatur}, {S.} {Inouye}, {H.-J.} {Miesner}, {J.} {Stenger}, and {W.} {Ketterle}, {Phys. Rev. Lett.}, \textbf{80}, {2027} ({1998})%
\bibitem{Ketterle:SpinDomainsInFroundStateOfBEC}%
  {J.} {Stenger}, {S.} {Inouye}, {D.} {Stamper-Kurn}, {H.-J.} {Miesner}, {A.} {Chikkatur}, and {W.} {Ketterle}, {Nature}, \textbf{396}, {345} ({1998})%
\bibitem{Stamper-Kurn:SpinTextures}%
  {M.} {Vengalattore}, {S.~R.} {Leslie}, {J.} {Guzman}, {D.~M.} {Stamper-Kurn}, {Phys. Rev. Lett.}, \textbf{100}, {170403} ({2008})%
\bibitem{Tsubota:DomainFormationInBEC}%
  {K.} {Kasamatsu} and {M.} {Tsubota}, {Phys. Rev. Lett.}, \textbf{93}, {100402} ({2004})%
\bibitem{Ho:PhaseDiagramOfSpinorBEC}%
  {C.~V.} {Ciobanu}, {S.-K.} {Yip}, {T.-L.} {Ho}, {Phys. Rev. A}, \textbf{61}, {033607} ({2000})%
\bibitem{Ho:GroundStateOfSpinorBEC}%
  {T.-L.} {Ho} and {S.~K.} {Yip}, {Phys. Rev. Lett.}, \textbf{84}, {4031} ({2000})%
\bibitem{Ueda:ExactEigenstatesOfSpinorBEC}%
  {M.} {Koashi} and {M.} {Ueda}, {Phys. Rev. Lett.}, \textbf{84}, {1066} ({2000})%
\bibitem{Wadati:SpinorSolitons}%
  {J.} {Ieda}, {T.} {Miyakawa}, and {M.} {Wadati}, {Journal of the Physical Society of Japan}, \textbf{73}, {2996} ( {2004})%
\bibitem{Wadati:DarkSolitons}%
  {M.} {Uchiyama}, {J.} {Ieda}, and {M.} {Wadati}, {Journal of the Physical Society of Japan}, \textbf{75}, {064002} ({2006})%
\bibitem{Malomed:SolitonsNonlinearModulation}%
  {L.} {Li}, {Z.} {Li}, {B.~A.} {Malomed}, {D.} {Mihalache}, and {W.~M.} {Liu}, {Phys. Rev. A}, \textbf{72}, {033611} ({2005})%
\bibitem{Wadati:BrightSpinorSolitons}%
  {J.} {Ieda}, {T.} {Miyakawa}, {M.} {Wadati}, {Phys. Rev. Lett.}, \textbf{93}, {194102} ({2004})%
\bibitem{Wadati:KdV}%
  T. Tsuchida and M. Wadati, Journal of the Physical Society of Japan, \textbf{67}, {1175} (1998)%
\bibitem{Ablowitz+Segur:Solitons+InverseScattering}%
  M. Ablowitz and H. Segur, \emph{Solitons and the Inverse Scattering Transform}, SIAM, Philadelphia, ({1981})%
\bibitem{MyPRL:Oscillatons}%
  P. Szankowski, M. Trippenbach, E. Infeld, and G. Rowlands, {Phys. Rev. Lett.}, \textbf{105}, {125302} ({2010})%
\bibitem{Landau+Lifshitz:vol3}%
  L. Landau and L. Lifshitz, \emph{Quantum Mechanics
  Non-Relativistic Theory, Third Edition: Volume 3}, Pergamon Press Ltd. (1977) pp. 73--74%
\bibitem{Whitaker+Watson:AcourseOfModernAnalysis}%
  E. {Whitaker} and G. Watson, \emph{A course of modern
  analysis}, Cambridge University Press, Cambridge (1996)
\bibitem{Morse+Feshbach:MethodsOfTheoreticalPhysics}%
  P. Morse and H. Feshbach, \emph{Methods of
  theoretical physics}, Vol.~{1}, {Mc Graw Hill, N.Y.} (1953)%
\end{thebibliography}

%

\end{document}